# Prediction and Inference: From Models and Data to Artificial Intelligence

**Luca Gammaitoni[1,2] · Angelo Vulpiani[3]**

**Abstract**
In this paper we present a discussion of the basic aspects of the well-known problem of prediction and inference in physics, with specific attention to the role of models, the use of data and the application of recent developments in artificial intelligence. By focussing in the time evolution of dynamic system, it is shown that main difficulties in predictions arise due to the presence of few factors as: the occurrence of chaotic dynamics, the existence of many variables with very different characteristic time-scales and the lack of an accurate understanding of the underlying physical phenomena. It is shown that a crucial role is assigned to the preliminary identification of the proper variables, their selection and the identification of an appropriate level of description (coarse-graining procedure). Moreover, it is discussed the relevance, even in modern practical issues, of old well-known fundamental results, like the Poincaré recurrence theorem, the Kac's lemma and the Richard's paradox.

## 1 Introduction

The aim of this paper is a general discussion of the basic aspects of prediction and inference; in particular the role of models at different scales, the importance of the proper level of description, the relevance of old (apparently very far) classical issues, as well as some remarks on inductivism and *Big Data*.

---

Luca Gammaitoni and Angelo Vulpiani have contributed equally to this work.

✉ Luca Gammaitoni
luca.gammaitoni@unipg.it

1 Dipartimento di Fisica e Geologia, Universitá degli Studi di Perugia, Via A. Pascoli, 06100 Perugia, Italy

2 NiPS Laboratory, Universitá degli Studi di Perugia, Via A. Pascoli, 06100 Perugia, Italy

3 Dipartimento di Fisica, Universitá La Sapienza di Roma, Piazzale Aldo Moro, 1, Roma, Italy

Generally speaking, prediction in physics is the activity aimed at forecasting a future event based on the use of models and theories. On the other hand, inference is the process by which, using data obtained from observation, the physical model is guessed. Traditionally, both prediction and inference are used in conjunction and constitute key aspects of our understanding and of the advancing of the knowledge in physics. The extent to which one is used versus the other, varied significantly along the history of physics. Just to present two very distinct approaches, we could mention the *strong reductionism* and the *naive inductivism* approaches.

The first approach is well represented here by the opinion of S. Weinberg [1]:

*Cold fronts are the way they are because of the properties of air and water vapour and so on which in turn are the way they are because of the principles of chemistry and physics.... We do not know the final laws of nature, but we know that they are not expressed in terms of cold fronts or thunderstorms.*

In this case the inference obtained from the observation of "cold fronts" does not provide any guess on the "final laws of nature", whatever they are. The opposite approach, the *naive inductivism*, is based completely on the sole inference from data, without any involvement of physical laws. It is, sometime, summarised in the slogan *forget the theory, now the data are enough*. The prophet of such a point of view has been, probably, C. Anderson, with his provocative paper *The Data Deluge Makes the Scientific Method Obsolete* [2], where the key idea is the following:

*Petabytes* (of data) *allow us to say: "Correlation is enough". Therefore we can stop looking for models. We can analyse the data without hypotheses about what it might show. We can throw the numbers into the biggest computing clusters the world has ever seen and let statistical algorithms find patterns where science cannot.*

There are strong evidences which suggest that both these approaches cannot work.

## 2 Prediction and Inference

The activity of producing predictions in physics is quite demanding, and at least two very different situations can be met, where inference plays a different role:

(A) We can be in a situation where we have a working theory (or, at least, a set of reliable models, where the relevant variable are identified);
(B) We have no working theory but only a large data set (often incomplete) of the past history of the system.

In the following we discuss these two quite different situations.

### 2.1 (A) The Prediction of the Future with a Working Theory

Let us start with the simplest case, where we deal with a physical system for which we know a set of variables $\mathbf{x} = (x_1, ..., x_N)$ that describe the phenomena we are interested in, and where we have a known evolution law for those variables. This law is usually provided by the solution of a differential equation. This is clearly an optimal condition, where we know both the *proper* variables **x** and the evolution law.

In order to understand the meaning of *proper* variables, let's see and example: the harmonic oscillator. For such a system the *proper* variables are represented by the position $x(t)$ and velocity $v(t)$ of the point mass, and the evolution law is represented by:

$$\frac{dx}{dt} = v \ , \ m\frac{dv}{dt} = -kx, \tag{1}$$

In such a system it is quite easy to predict the future from $x(0)$ and $v(0)$:

$$x(t) = x(0)cos(\omega t) + \frac{v(0)}{\omega}sin(\omega t) \ , \ \omega = \sqrt{\frac{k}{m}}. \tag{2}$$

However, even if this example represents a very popular case, we should admit that this situation is not so common in any real physics problem. In fact, in most cases, even if the evolution law is known, in general it is not possible to find explicitly a mathematical expression (analytic solution in closed form) for it. However, with the present existence of powerful numerical computing methods, one could think that, the lack of an explicit mathematical expression is not a serious limitation for the goal of predicting the future evolution of the system. Indeed, this was also the hope of J. von Neumann, who believed that, using large computers and efficient numerical methods, it was possible to predict, and even to control, the weather and the climate. Unfortunately, in his arguments he did not take into account the possibility to be faced with the chaos phenomenon, whose existence had been discovered during the study of well known "three body problem", by Poincaré [3].

To briefly recall the role of chaos in the evolution of a dynamical system, we can simply cite the (in)famous butterfly effect, discussed by Lorenz: two particular weather situations differing by as little as the immediate influence of a single butterfly wing flap, say $|\delta x(0)|$, will generally, after sufficient time, evolve into two situations differing by as much as the presence of a tornado. After the (re)discovery of chaos in the 1960 s, it is now well clear that determinism does not imply accurate prevision at large time, and in addition even low dimensional "simple" systems can be chaotic and can show highly non-trivial behaviours.

We can formalise the main aspect of chaos, in terms of the Lyapunov exponent $\lambda$ [3] In few words: in a deterministic chaotic system, for small values of $|\delta \mathbf{x}(t)|$ we have

$$|\delta \mathbf{x}(t)| \sim e^{\lambda t}|\delta \mathbf{x}(0)| \tag{3}$$

with $\lambda > 0$. The relevance of the Lyapunov exponent for the prediction is quite transparent: it is possible to perform a prediction with a tolerance $\Delta$ only until a time $T_p$ which depends on $\lambda$ and (weakly) on the initial uncertainty $|\delta \mathbf{x}(0)|$

$$T_p \sim \frac{1}{\lambda} \ln\left(\frac{\Delta}{|\delta \mathbf{x}(0)|}\right). \tag{4}$$

We do not deal here in the mathematical treatment of the problem and the necessity of a generalisation to the case of non infinitesimal $|\delta \mathbf{x}(t)|$. A detailed discussion can be easily found in the literature [4].

### 2.2 (B) The Prediction of the Future Without a Working Theory

Let's now address case (B) where, at difference with the previous case, we do not have a working theory. In this case, we can identify two distinct situations:

(B1) We miss the knowledge of the evolution law but we know the *proper* variables **x**.
(B2) We miss the knowledge of the evolution law and we do not know the *proper* variables **x**, but we have only time series of some characteristic observables.

#### 2.2.1 (B1) We Know the *Proper* Variables

Consider now the case where in our prediction problem we know that the state of the system is represented by a vector **x** that represents the *proper* variable. Let's suppose that we know the past of the system, i.e. a time series $(\mathbf{x}_1, \mathbf{x}_2, ...., \mathbf{x}_M)$, where $\mathbf{x}_j = \mathbf{x}(j\Delta t)$, and we want to predict the future, i.e. $\mathbf{x}_{M+t}$ for $t > 0$.

Assuming the the system is deterministic, in order to predict the future it is enough to look into the past for an *analog* i.e. a vector $\mathbf{x}_k$ with $k < M$ such that $|\mathbf{x}_k - \mathbf{x}_M| < \epsilon$. Therefore, we can predict the future at times $M + t > M$:

$$\mathbf{x}_{M+t} \simeq \mathbf{x}_{k+t}. \tag{5}$$

Of course, in the presence of chaos, as we learned above, this procedure can be used only for small values of $t$. However difficult, as we will soon see, this is not the most serious limitation that we are faced with.

In fact, in order to succeed with this method, it is necessary to find the *analog*, as defined above. How hard it could be?

In the 1960 s Lorenz tried to use the meteorological charts of the past to perform a weather forecasting [5]. He soon realised that the application of this method was more problematic than what was expected and, finally, that it did not work at all. The failure of the method, as well understood by Lorenz himself, is due to the fact that there is a high probability that no *analog* good enough will be found within the whole recorded history of the atmosphere.

This was long time ago. Presently, there is no doubt that we are in a much better position compared to Lorentz. As a matter of fact we are in the age of *Big Data*, can we hope to success using just data where Lorenz did not?

In order to answer this question we need to recall an apparently old an distant topic, meaning the *Poincaré recurrence theorem* [3]: *In a deterministic system with a bounded phase space, after a certain time, the system must come close to its initial state.*

Such a theorem had a great relevance in the heated debate that took place at the end of the 19-th century on the irreversibility problem[6], between Boltzmann and Zermelo. Boltzmann suggested that, in a system with $N \gg 1$ particles, the return time can be pretty large and, specifically, as large as

$$T_R \sim \tau_0 C^N \tag{6}$$

where $\tau_0$ is a characteristic time and $C > 1$. In a macroscopic system $N$ is very large (at least $10^{20}$) therefore $T_R$ can easily be larger than the age of the universe.

The intuition of Boltzmann has been subsequently formalised in the Kac Lemma [7]:

*In an ergodic system the average return time in a region $\langle \tau(A) \rangle$ in a set A is $\langle \tau(A) \rangle = \tau_0/P(A)$ where $P(A)$ is the probability to be in A.*

As can be easily shown, such a result has a rather negative consequence for the forecasting task. In fact, let's consider a region $A$ of linear sizes $O(\epsilon)$, therefore $P(A) \sim (\epsilon/L)^D$, where $L$ is the excursion of each component of the vector **x** and $D$ the attractor's dimension, thus

$$\langle \tau(A) \rangle \sim \tau_0 (L/\epsilon)^D \tag{7}$$

Therefore, in order to find an analog, the time series length M, must be at least of the same order of the recurrence time:

$$M_{min} \sim \tau_0 (L/\epsilon)^D / \Delta t \tag{8}$$

as consequence, even with a limited precision, say 5%, i.e. $L/\epsilon = 20$, if $D$ is large, say $D = 6$ or $D = 7$, it is practically impossible to find an *analog*.

An interesting case of the use of the *analog* approach is represented by the prediction of the phenomenon of tides. Already in the 19th century, there existed efficient empirical methods to compile numerical tables of tides in any location, using a record of past tides. Lord Kelvin and George Darwin (Charles's son) showed that water levels can be well predicted by a limited number of harmonics (say 10 or 20), determining the Fourier coefficients from the past time data at the location of interest. Kelvin, and his brother, built a tide-predicting machine: a special purpose mechanical computer made of gears and pulleys [8].

Lord Kelvin and George Darwin were surely very smart, but also rather lucky: actually tides are chaotic, nevertheless their prediction from past records was a relatively easy, successful task. Now, a posteriori, we understand the reason of the success of the method: although chaotic, the dynamics is such that the attractor dimension is quite low $O(3-4)$ [9]. On the contrary, Lorenz although very smart, was rather unlucky; actually he had no chance to find an *analog*: for the atmosphere, being $D$ in this case, not small at all. Most likely, the dimension $D$ here is $O(10^3 - 10^4)$.

### 2.2.2 (B2) We do not Know the *Proper* Variables

Finally, we find ourselves in a quite common situation: the case where we do not know the proper variables but just a time series $\{u_1, .., u_M\}$, where $u_j$ is an observable sampled at the discrete times $j\Delta t$. It is natural to wonder whether, in such a case, it is possible to guess the underlying dynamics from the data.

The mathematician F. Takens had been able to give a wonderful contribution to this problem [10]:

*From the analysis of a time series $\{u_1, .., u_M\}$, where $u_j$ is an observable sampled at the discrete times $j\Delta t$, it is possible (if we know that the system is deterministic and is described by a finite dimensional vector, and M is large enough) to determine the proper variable* **x**,*and then determine the main features of the dynamics, e.g. the Lyapunov exponent and D.*

Unfortunately in the practical world, Takens's result is not a panacea. As a matter of fact, there are rather severe limitations in practical cases. Most importantly, there is the necessary preliminary knowledge of the deterministic character of the dynamics. In addition, as consequence of the Kac's lemma, the protocol fails if the dimension of the attractor is large enough (say larger than $D = 5$ or $D = 6$) [9].

In spite of the many delusions that, due to the technical severe limitations to a purely inductive approach, appeared soon after the initial enthusiasm (the happy chaotic years 1980 s-1990 s), there are presently still relevant attempts of pursuing the naive baconian dream of a science without equations. Sometimes these even appear in prestigious scientific journals [11].

## 3 A Relevant Case: Weather Prediction

Let us briefly discuss the history of weather forecasting. Indeed, such a topic is rather useful for the understanding of the role of theory, models and computer [12, 13].

It is well known that the dynamics of the atmosphere follows to the laws of hydrodynamics and thermodynamics. Thus, from the knowledge of the present state of the atmosphere (i.e. the fields describing velocity **u**, density $\rho$, pressure $p$ etc) we can solve a set of partial differential equations to obtain, at least in principle, a description of the future state of the atmosphere, i.e. the weather forecast. In this case we are, clearly, in the relatively simple case where we know the evolution laws and the *proper* variables.

It is known that the relevant equations cannot be solved analytically and we may want to resort to a numerical solution.

The modern pioneer of the weather forecasting was undoubtedly L.F. Richardson [12, 14]. It is well known that in his first attempt in producing weather forecasting reports, he used as initial conditions the weather charts compiled with the observations taken in Northern Europe at 4 a.m. on 20 May 1910 during an international balloon day. After a long, taxing and wearisome numerical work, Richardson finally obtained a six-hour forecast, but the result was rather disappointing.

Paradoxically, the reason of this failure was the fact that Richardson was too accurate. In computing the time evolution, he used the "true" equations, i.e. those able to describe all the fluid features and associated phenomena, including the so-called "fast phenomena", e.g. the waves in the fluids. Unfortunately, such a detailed description had a very negative impact on the numerical computations. The way around this limitation was found, few years later, by Charney, von Neumann and colleagues, within the Meteorological Project at the Institute for Advanced Study, Princeton (1940 s- 1950 s) [15]. Instead of the use of the "true" equations, derived from first principles, Charney and von Neumann adopted another approach. This was based on the introduction of three new "ingredients", all far from trivial. Namely, the use of simplified "effective equations", fast numerical algorithms and computers for numerical calculations.

Let us briefly comment the relevance of the introduction of the "effective equations". Almost all the interesting dynamic problems in science and engineering are characterised by the presence of more than one significant scale, i.e. there is a variety of degrees of freedom, each characterised by different time scale. A relevant example is represented by the climate dynamics, where the characteristic times of the involved processes vary from days (for the atmosphere) to $O(10^3 - 10^4)yr$ (for the deep ocean and ice shields) [16].

The necessity of treating the "slow dynamics" in terms of "effective equations" is both practical (even modern supercomputers are not able to simulate all the relevant scales involved in certain difficult problems) and conceptual: "effective equations" represent a less accurate and detailed description of the phenomena that are able to catch some general features and to reveal dominant ingredients, that otherwise remain hidden in a detailed description.

Moreover, the advantages of the effective equations are relevant from a computational point of view: it is possible to use larger $\Delta t$ and $\Delta x$ in the numerical integration and their description of the slow dynamics, make it possible to detect the most important factors, which on the contrary remain hidden in the detailed description given by the original equations[17].

Let us stress the fact that often the "effective equations" are not mere approximations of the original equations because, typically, emergent features appear [17].

## 4 Building Models from the Sole Data: Fundamental Difficulties

Quite recently, some authors started to believe that, since the recent developments on Data Science and Artificial Intelligence, physics may be moving into a period where equations do not play the central role in describing dynamic systems that they have played in the last 300 years [11, 18].

Such an idea is not completely new, e.g. in the '80 s, some researchers in the field of artificial intelligence (AI) devised BACON, a computer program to automatise scientific discoveries. Apparently, BACON was able to "discover" the third Kepler's law [19].

Let us look at the details of the procedure used by BACON, as compared to the one used by Kepler: BACON used as input the numerical values of distance from the Sun, *d*, and revolution period, *p*, of planets. The program, then, discovered that $d^3$ is proportional to $p^2$.

At difference, Kepler did not have the listed values of *d* and *p* but sets of raw observables consisting in a list of planetary positions as seen from the Earth at different times. In his discovery, Kepler had to guess the "right" variables *d* and *p*, while he was guided by strong beliefs in mathematical harmonies as well as in the (at that time) controversial theory of Copernicus.

A similar situation has recently arisen in the framework of gravitational wave detection. After the initial discovery made in 2015 [20] the existing detectors have been accumulating very large amount of data. The detection process is performed by searching for waveforms hidden inside usually very noisy time series. After scanning in dept the data for candidates taken from the catalog of expected waveforms, someone suggested that the search activity could be extended to unknown waveforms, by using Artificial Intelligence (i.e. deep learning) techniques.

Also in this case, the problem is not completely new. In fact, it can be rephrased into the more general *search for something without knowing exactly what you are searching for* kind of problem. Now, there a number of reasons why this is task that is doomed to fail. The most interesting explanation can be understood by recognising that this problem is no different by the so called "Babel library" problem.

### 4.1 Why it is Difficult to Find Something If You do not Know What You are Looking For

In the famous short novel by Jorge Luis Borges entitled *The library of Babel*, the author describes an imaginary, vast library, composed of a huge number of hexagonal rooms, all the same, connected by corridors. As Borges explains: *Each wall of each hexagon corresponds to five shelves; each shelf contains thirty-two books of uniform format; each book is four hundred and ten pages; each page, of forty lines; each a row, of forty letters of black colour* [21].

As will soon be discovered in the story, each book is composed of a random sequence of symbols (22 letters, plus the comma, the point and the space). The library contains all the possible combinations of these symbols and, thus, contains all the writable books that satisfy the conditions of length above expressed. To use Borges' poetic language, the library's books describe everything: *the meticulous history of the future, the autobiographies of the archangels, the faithful catalogue of the Library, thousands and thousands of false catalogues, the demonstration of the falsity of these catalogues, the demonstration of the falsity of the authentic catalogue, the Gnostic gospel of Basilides, the commentary of this gospel, the commentary of this gospel, the truthful account of your death, the translation of each book in all languages, the interpolations of each book in all the books*.

The problem facing the visitor of the library is clearly to decipher the books because a book taken at chance from the shelf appears as a sequence of symbols randomly assembled to compose meaningless words. Apparently meaningless but, perhaps, expressing the story of our life or a prophecy or even the final equation of Physics, in another unknown and mysterious language.

A book from the library of Babel, definitely looks just like a series of data gathered from the experiment on gravitational waves detection, provided that you assign a number to each symbol. It is apparently a sequence of random numbers but you never know if, hidden within the ocean of noise, there is some promising signal. Therefore, looking for a new signal in the experimental data series would not be very different from looking for a sensible and interesting expression, in one of the library's books. And here comes the beauty: once this analogy is established (a book as a bunch of data) the Library of Babel really looks like the Big Data paradise. It contains all the information of potential interest to us, the problem is "just" extracting this information: how hard can it be?

Unfortunately, hard to the point of being impossible. In fact, even if the number of books in the *Library of Babel* is very large but still finite, the number of meanings that can be attributed to the content of those books is not. Technically this is called "Richard's paradox" and is part of a family of results that led the logical-mathematician Gödel to formulate his famous incompleteness theorems [22]. An infinite number of meanings potentially correspond to a finite string of characters, or in our context, a certain set of data constitutes the potential answer to an infinite number of physical questions. Without knowing the question, the answer risks being meaningless. Without using a proper dictionary, the string of characters has no meaning.

Out of metaphor, the construction of the scientific model (the dictionary in our example), is crucial for the interpretation of the experimental data. Without the laborious, complicated and "dirty" work of the scientist who mixes intuition and induction, genius and creativity, manages to advance hypotheses that will then be denied or confirmed by data, there is no production of true knowledge. To paraphrase the great French scientist Poincareé, we could say that the scientist must make order: *science is made with data as a house is made with bricks, but the accumulation of data is not science any more than a pile of bricks is a house*.

## 5 Conclusions

Perhaps is too early to decide whether Artificial Intelligence methods are able replace the traditional creative approach to scientific discoveries, even if dealing with large quantities of raw data, some imitations appear well evident. To summarise some interesting caveat, let us list some facts:

(a) Even in the (lucky) case we know the proper variables $\mathbf{x}_t$, as consequence of the Kac's lemma, if the dimension is larger that $D = 5$ or $D = 6$ it is pretty impossible to find analogs [9].
(b) Typically we do not know the proper variables; such rather serious difficulty is well known, for instance in statistical physics, let us cite an important caveat from Onsager and Machlup: *How do you know you have taken enough variables, for it to be Markovian?* [23]
(c) In generic stochastic systems the Takens protocol for the phase space reconstruct does not hold [24].
(d) The problem of discriminating equilibrium from non equilibrium has no answer using a series of scalar with a Gaussian statistics: observing a single variable (or a linear combination of variables) results in a one-dimensional process is always indistinguishable from an equilibrium one [25].

In a nutshell we can say that the main difficulties in the predictions are the presence of chaos, the many variables with very different characteristic times and the lack of accurate understanding of the underlying phenomena.

It is important to keep in mind that old topics can be relevant even in modern practical issues: e.g. the Poincaré recurrence theorem, and Kac's lemma. Moreover, the attempts to build models just from data have fundamental theoretical limits (Richard's paradox) and severe practical limits (large dimensionality of the problem).

**Author Contributions**  All the authors contributed equally.

**Funding**  Open access funding provided by Università degli Studi di Perugia within the CRUI-CARE Agreement.

**Data Availability**  No datasets were generated or analysed during the current study.

**Declarations**

**Conflict of interest**  The authors declare no conflict of interest.